\documentclass[conference]{IEEEtran}
\IEEEoverridecommandlockouts
\usepackage{graphicx,cite,calc,xcolor,subfigmat,amssymb,amsmath,mathrsfs,dsfont,hyperref,epstopdf}
\usepackage[normalem]{ulem}
\usepackage{algorithm}
\usepackage[switch]{lineno}

\usepackage{algpseudocode}
\usepackage{algorithmicx}

\usepackage{dsfont}
\usepackage{bbm}
\usepackage{soul}
\usepackage{xcolor}
\usepackage{amsmath}

\usepackage{float}
\usepackage[T1]{fontenc}
\usepackage{soul} 
\usepackage{xcolor}
\usepackage{optidef}
\usepackage{listings}
\usepackage{enumitem}
\usepackage{hyperref}
\makeatletter
\newcommand*{\rom}[1]{\expandafter\@slowromancap\romannumeral #1@}
\makeatother
\usepackage {amsmath}
\definecolor{highlightcolor}{RGB}{255,255,0}

\usepackage{newtxtext,newtxmath}
\title{Novel Methods for Load Estimation in Cell Switching in HAPS-Assisted Sustainable 6G Networks}

\begin{document}

 \author{\rm{Maryam Salamatmoghadasi}, \rm{Metin Ozturk}, \rm{Halim Yanikomeroglu}
 \thanks{This research has been sponsored in part by the NSERC Create program entitled TrustCAV and in part by The Scientific and Technological Research Council of Türkiye (TUBITAK) under the TUBITAK-2219 program.
 
All the authors are with Non-Terrestrial Networks Lab, Department of Systems and Computer Engineering, Carleton University, Ottawa, ON K1S5B6, Canada. Metin Ozturk is also with Electrical and Electronics Engineering, Ankara Yildirim Beyazit University, Ankara, 06010, Turkiye. emails: \texttt{maryamsalamatmoghad@cmail.carleton.ca, metin.ozturk@aybu.edu.tr, halim@sce.carleton.ca}.}}

\maketitle
\begin{abstract}
In the evolving landscape of vertical heterogeneous networks (vHetNets), the practice of cell switching—particularly for small base stations (SBSs)—faces a significant challenge due to the lack of accurate data on the traffic load of sleeping SBSs. 
This information gap is crucial as it hinders the feasibility and applicability of existing power consumption optimization methods; however, the studies in the literature predominantly assume perfect knowledge about the traffic load of sleeping SBSs. 
Addressing this critical issue, our study introduces innovative methodologies for estimating the traffic load of sleeping SBSs in a vHetNet including the integration of a high altitude platform (HAPS) as a super macro base station (SMBS) into the terrestrial network. 
We propose three distinct spatial interpolation-based estimation schemes: clustering-based, distance-based, and random neighboring selection. 
Employing a real data set for empirical validations, we compare the estimation performance of the developed traffic load estimation schemes and assess the impact of estimation errors. 
Our findings demonstrate that accurate estimation of sleeping SBSs' traffic loads is essential for making network power consumption optimization methods both feasible and applicable in vHetNets.
\end{abstract}
\section{Introduction}
\indent The rapid escalation of mobile traffic demand, spurred by the proliferation of multimedia content and the internet of things (IoT), necessitates the densification of radio access networks (RANs). 
With the advent of the sixth generation of cellular networks (6G) networks, which are expected to support an unprecedented number of devices per square kilometer, the quest for enhanced connectivity becomes increasingly vital. However, this expansion is not without challenges, most notably the significant rise in energy consumption within RANs. 
Base stations (BSs), in particular, are major contributors to this energy usage in cellular networks, as highlighted in \cite{3333333}.

Addressing this challenge necessitates a re-evaluation of current operational practices. Traditionally, the practice of keeping all BSs—especially small BSs (SBSs)—continuously active, irrespective of actual user demand, leads to notable energy wastage~\cite{4444444}. This problem is exacerbated by the inherently fluctuating nature of cellular network traffic, which varies both temporally and spatially. A strategic solution lies in the deactivation of BSs, i.e., cell/BS switching off~\cite{3333333}. 
This method, which involves switching off SBSs or putting them into sleep mode during periods of low activity, offers a feasible strategy for enhancing energy efficiency and network sustainability. 
However, the implementation of effective cell switching approaches faces a significant hurdle: the lack of precise traffic load information for sleeping SBSs. 
This absence of data renders current methods for optimizing power consumption impractical and unworkable, since the majority of the works in the literature assume perfect knowledge on the traffic loads of sleeping BSs~\cite{10304250,HETS2023P,ELAA2022JR,EOMK2017JR,Metin_VFA_CellSwitch}.
The key to making cell switching strategies viable is addressing the challenge of estimating the traffic loads of sleeping SBSs. 
Tackling this issue not only bridges the gap between theory and practice but also unlocks the full potential of cell switching approaches for significantly improving sustainability in vertical heterogeneous networks (vHetNets)~\cite{8833522}.

The literature on cell switching extensively investigates policies for more efficient BS deactivation, with the primary goal of reducing overall network energy consumption. 
For example, the research presented in~\cite{10304250} explored a vHetNet model, which incorporated a high altitude platform station (HAPS) functioning as a super macro base station (SMBS), in addition to a macro BS (MBS) and several SBSs. 
The focus was on optimizing the sleep mode management of SBSs and leveraging the capabilities of HAPS-SMBS to reduce energy consumption in vHetNets, while ensuring the maintenance of user quality-of-service (QoS). 
However, a pivotal limitation in this study is the assumption of complete knowledge of sleeping SBSs' traffic loads—an assumption that necessitates accurate estimation in reality.
In a similar vein, the study presented in~\cite{HETS2023P} explored the cell switching conundrum, particularly in the presence of HAPS-SMBS. 
The authors focused on offloading the traffic loads from deactivated BSs to HAPS-SMBS. 
A notable strategy employed in this investigation involves the use of a sorting algorithm that prioritizes BSs with relatively lower traffic loads for switching off. The reliance of their model on traffic load data highlights the existing issue of accurate load estimation for these studies.

In \cite{ELAA2022JR}, a tiered sleep mode system for BSs was proposed, where sleep depth varied with the number of devices in sleep mode. 
The authors implemented a decentralized control, allowing each BS to independently manage its sleep strategy, thus keeping the system scalable and computationally efficient. 
A simplified $Q$-learning algorithm was used, although the issue of accurate traffic load estimation for sleeping cells remained unaddressed.
Another interesting perspective was proposed in~\cite{ENHF2023JR} and \cite{EOMK2017JR}, where the authors implemented a genetic algorithm to optimize energy savings by switching off BSs within a heterogeneous network (HetNet) environment. 
This approach primarily revolved around user association, following which BSs are deactivated in accordance with a deterministic algorithm. 
The study also considered the power consumption implications of transitioning between different BS states, namely ON, OFF, and sleep, and integrated the control data separated architecture (CDSA) into its framework. 
Similar to previous studies, the issue of accurately estimating the traffic load of sleeping cells is also a challenge here.
Moreover, the authors in~\cite{Metin_VFA_CellSwitch} proposed an innovative approach to cell switching in ultra dense networks by employing a value function approximation (VFA)-based reinforcement learning (RL) algorithm. 
The results demonstrated potential energy savings while maintaining the QoS, suggesting a scalable and less complex solution compared to exhaustive search methods.
However, the challenge of traffic load estimation for sleeping BSs remains unaddressed, limiting the practical application of these studies and their contribution towards achieving sustainability goals of 6G networks as mentioned in the 6G framework of International Telecommunication Union~(ITU)~\cite{itu_vision_june_23}.

Addressing this gap, our work stands as a pioneering effort to investigate and evaluate various spatial interpolation methods for estimating the traffic load of sleeping cells. 
Without properly addressing the novel cell load estimation problem presented in this work, the applicability and feasibility of the existing literature about cell switching approached are always questionable.
In this regard, we conduct a comparative analysis of the estimated values against their actual counterparts and evaluate the implications of any discrepancies on the network's power consumption within a vHetNet, as analyzed in~\cite{10304250}.
Our study makes significant contributions in the following areas:
\begin{itemize}
\item Introducing and investigating the sleeping SBSs' load estimation problem for cell switching and traffic offloading in a vHetNet.
\item Developing three innovative spatial interpolation algorithms---$k$-means clustering, distance-based, and random neighboring selection---for traffic load estimation. 
\item Utilizing a real data set~\cite{DVN/EGZHFV_2015} to enhance the realism and reliability of our approach.
\item Evaluating the impact of traffic load estimation errors on the network power consumption outcomes studied in~\cite{10304250}.
\end{itemize}

The rest of the paper is organized as follows: Section~\ref{sec:model} presents the system model, including network and power consumption models as well as data processing, while the problem formulation is elaborated in Section~\ref{sec:problem}.
The proposed cell load estimation techniques are introduced in Section~\ref{sec:method}, followed by discussing the obtained results in Section~\ref{sec:performance}.
Lastly, Section~\ref{sec:conclusion} concludes the paper.

\section{System Model}\label{sec:model}
In this section, we present an overview of the network model, including its components and characteristics. We also detail the metrics used to measure network power consumption and describe the parameters of the data set employed for empirical validation.
\subsection{Network Model}
\begin{figure}[t]
\centerline{\includegraphics[width = \columnwidth]{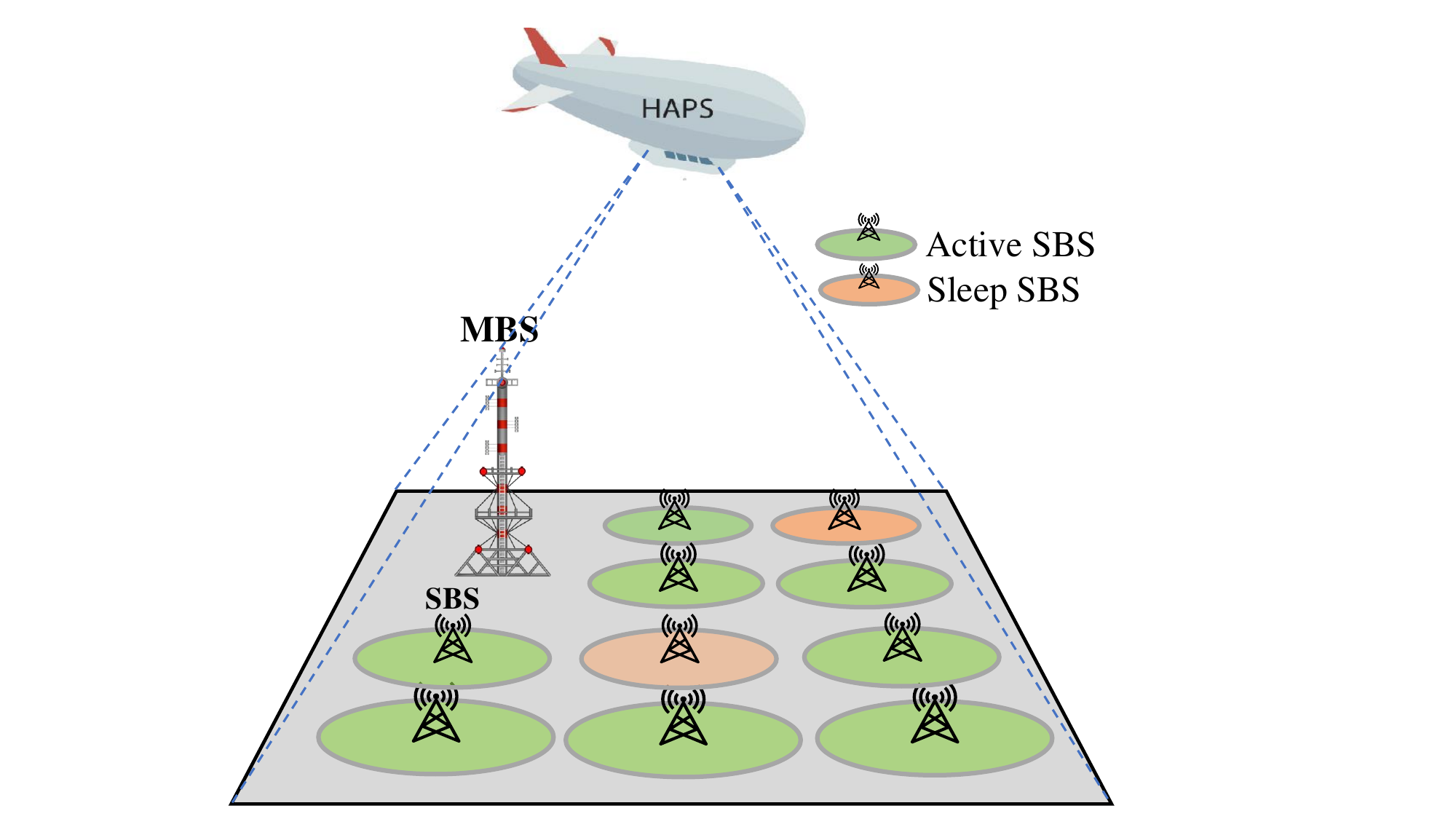}}
\caption{A vHetNet model with an MBS, multiple SBSs, and a HAPS-SMBS.}
\label{fig-1}
\end{figure}

Our study examines a vHetNet as illustrated in Fig. \ref{fig-1}, which consists of a macro cell (MC), encompassing a MBS and a cluster of SBSs within its coverage area. 
Additionally, a HAPS-SMBS is integrated into the network, potentially serving multiple MCs. 
The primary function of SBSs is to deliver data services and address user-specific requirements, while MBS and HAPS-SMBS are tasked with providing consistent network coverage and managing control signals, along with offering data services.
A key role of the HAPS-SMBS in this architecture is to coordinate the offloading of traffic from SBSs, particularly during low-traffic periods. This is facilitated by HAPS's higher capacity and extensive line of sight, enabling efficient monitoring of SBSs' traffic loads~\cite{9380673}. 
Based on this monitoring, strategic decisions are made regarding which SBSs to switch off, enhancing the network's operational efficiency. 
These decisions take into account the current traffic capacity of both the MBS and HAPS-SMBS to ensure efficient network operation.

\subsection{Network Power Consumption}
The power consumption of each BS in the network is calculated-based on the energy-aware radio and network technologies (EARTH) power consumption model~\cite{6056691}. 
For the $i$-th BS, the power consumption at any given time, denoted as $P_i$, is expressed as \cite{7925662}
\begin{equation}
P_i  = \left\{ {\begin{array}{*{20}c}
   {\begin{array}{*{20}c}
   {P_{\text{O},i}  + \eta _i \lambda _i P_{\text{T},i} }, & {0 < \lambda _i  < 1,}  
\end{array}}  \\
   {\begin{array}{*{20}c}
   {P_{\text{S},i} },  \; \;   \; \; \; \; \; \; \;\; \; \; \; \; \; \; \; \; \; \; \; \; \; & {\lambda _i  = 0},  \\
\end{array}}  \\
\end{array}} \right.
\label{eq1}
\end{equation}
where $P_{\text{O},i}$ represents the operational circuit power consumption, $\eta_i$ is the power amplifier efficiency, $\lambda_i$ is the load factor, $P_{\text{T},i}$ is the transmit power, and $P_{\text{S},i}$ is the sleep circuit power consumption. 
The total instantaneous power consumption of the network, denoted as $P$, is given by
\begin{equation}
P = P_\text{H}  + P_\text{M}  + \sum\limits_{j = 1}^s {P_j }, 
\label{eq2}
\end{equation}
where $P_\text{H}$ and $P_\text{M}$ denote the power consumption of HAPS-SMBS and MBS at any given moment, respectively, which are calculated based on the $(0 < \lambda _i  < 1)$ case in \eqref{eq1} as HAPS-SMBS and MBS are always active in our modeling. 
Meanwhile, $P_j$ represents the power consumption of the SBSs, which is determined using both $\lambda _i$ cases of \eqref{eq1}, depending on their operational status. 
The variable $s$ signifies the total number of SBSs within the network.

\subsection{Data Set and Data Processing}
To calculate the power consumption using~\eqref{eq1}, the load factor $\lambda_i$ for each BS is required. 
For this purpose, we utilize a real call detail records (CDR) data set provided by Telecom Italia~\cite{DVN/EGZHFV_2015}. 
The data set divides the city of Milan into 10,000 square-shaped grids, each measuring $235\times 235$ meters. Within each grid, levels of user activity—including calls, text messages, and internet usage—were recorded at 10-minute intervals over two months (November and December 2013). In processing this data, we first amalgamate these distinct activity types into a singular measure. Then, for each SBS, a grid is randomly selected, and the combined activity levels from that grid are normalized to represent the traffic load for the corresponding cell.

\section{Problem Formulation}\label{sec:problem}
This section outlines the power consumption optimization problem in vHetNets and addresses the impact of traffic load estimation errors on the optimization policy.

\subsection{vHetNet Power Consumption Optimization Problem}
Our objective is to minimize the total power consumption of the vHetNet, as defined in \eqref{eq2}. We achieve this by identifying the most efficient power-saving state. The state vector ${\rm \Delta = \{ 
\delta _1 ,\delta _2 ,...,\delta _s \} } $ represents the status (ON or OFF) of each SBS at time $t$, where ${\rm \delta _j  \in \left\{ {0,1} \right\}}$ indicates the state of the $j$-th SBS (0 for OFF, 1 for ON). It is assumed that MBS and HAPS-SMBS are always ON ($\delta _\text{M} = \delta _\text{H} = 1$). 
For an SBS, when ${\rm \delta _j }$ changes from 1 to 0 at time $ {t} $, either MBS or HAPS-SMBS could allocate its traffic.
On the other hand, when ${\rm \delta _j }$ changes from 0 to 1, MBS or HAPS-SMBS offload some traffic to the $j$-th ${\rm SBS}$. 
The optimization problem can be formulated as

 \begin{IEEEeqnarray*}{lcl}\label{eq:P1}
    &\underset{\mathbf{\Delta}}{\text{minimize}}\,\, & ~P(\Delta ) \,  \IEEEyesnumber \IEEEyessubnumber* \label{eq:P1_Obj}\\
    &\text{s.t.} & {\lambda _\text{M}}{\le 1,} \label{eq:P1_const1}\\
    && {\lambda _\text{H}}{\le 1,} \label{eq:P1_const2}\\
    && {{\delta _j \in \{ 0,1\}, }}  {{\; j = 1,...,s}}, \label{eq:P1_const3}
\end{IEEEeqnarray*}
where
\begin{equation}
\begin{aligned}
P(\Delta )&={(P_{\text{O,H}}  + \eta _\text{H} \lambda _\text{H} } P_{\text{T,H}} )+{(P_{\text{O,M}}  + \eta _\text{M} \lambda _\text{M} } P_{\text{T,M}})\\
 &\hspace{-2em}\;\;\;\;\;\;\;+(\sum\limits_{j = 1}^s {(P_{\text{O},j}  + \eta _j \lambda _j  P_{\text{T},j} )\delta _j +P_{\text{S},j}(1-\delta _j))}.
\end{aligned}
\label{eq4}
\end{equation}
In the constraints \eqref{eq:P1_const1} and \eqref{eq:P1_const2}, $\lambda_\text{M}=[0,1]$ and $\lambda_\text{H}=[0,1]$ are defined as positive real numbers, $\mathbb{R}^+$. The constraints ensure that the operational capacities of both the MBS and HAPS-SMBS are never exceeded, thereby upholding the QoS requirements. Importantly, the optimal state vector, ${\Delta _{\text{opt}}}$, minimizes ${P(\Delta )}$, which is a function of the load factors, ${\lambda _j}$, of SBSs. 
Therefore, any errors in estimating these load factors can alter the optimal state vector, impacting the total power consumption of the vHetNet.

\subsection{Estimation of Sleeping SBS Traffic Load}
The optimization problem defined in \eqref{eq:P1} aims to minimize the vHetNet's total power consumption while adhering to specific constraints.
The decision to choose the best state vector, ${\Delta _{\text{opt}}}$, on the traffic loads of the SBSs, ${\lambda _j}$. 
A critical question arises: How can the traffic load of a sleeping SBS, which was previously set to sleep mode, be known for determining its status in the next time slot? 
In other words, obtaining ${\lambda _j}$ for sleeping SBSs is a problematic aspect, as existing literature on cell switching often assumes perfect knowledge of these traffic loads \cite{10304250}--\cite{Metin_VFA_CellSwitch}.
This assumption is impractical and unrealistic, as traffic loads should be estimated, inevitably introducing estimation errors. In the subsequent section, we propose three different methods to estimate the traffic load of sleeping SBSs, thereby addressing this critical gap in the field.

\section{Proposed Load Estimation Scheme}\label{sec:method}
This paper introduces novel methodologies for estimating the traffic load of sleeping BSs, a critical factor in optimizing network power consumption. In this regard, we propose three distinct spatial interpolation approaches, which will be elaborated in the following paragraphs, that are used to estimate the traffic loads of the sleeping SBSs. More specifically, the base methodology is to use spatial interpolation to estimate the traffic loads, and the three approaches we develop are utilized to select the cells (i.e., cell selection approaches) that would be involved in this interpolation process.

\subsection{Clustering-Based Traffic Load Estimation}
The first approach involves clustering SBSs based on their traffic loads and estimating the traffic load of a sleeping SBS using the average load of active SBSs within the same cluster. 
We employ the $k$-means algorithm, an unsupervised machine learning technique, for clustering. 
The number of clusters, a crucial hyper-parameter, is determined using the elbow method \cite{article}, which
assesses different cluster numbers by calculating the sum of squared errors (SSE) between data points and centroids for each potential cluster count. The SSE is given by \cite{article}
\begin{equation} 
SSE = \sum\limits_{k = 1}^K {\sum\limits_{x_m\in C_k} {(x_m - c_k )} } ^2. 
\label{eq6}
\end{equation}
where $K$ represents the optimal number of clusters, $c_k$ the centroid of each cluster $C_k$, and $x_m$ each sample in $C_k$. The optimal cluster number is identified at the point where the SSE curve forms an "elbow" before flattening.
\subsubsection{Multi-level Clustering-Based Traffic Load Estimation}
The multi-level clustering (MLC) approach involves repeated clustering of SBSs based on their traffic loads to estimate the traffic load of offloaded SBSs. This method employs the elbow method to determine the optimum number of clusters, followed by the application of the $k$-means algorithm for clustering. The traffic load of a sleeping SBS is then estimated based on the average load of active SBSs in the same cluster. This clustering process, which is detailed in Algorithm \ref{alg1}, is recursively applied to form multi-level clusters, thereby enhancing the accuracy of the estimations. This iterative approach, as outlined in Algorithm \ref{alg1}, ensures progressively refined clustering with each layer, leading to more precise traffic load estimations for sleeping SBSs.

\begin{algorithm}
\caption{Multi-Level Clustering (MLC) using $k$-means}
\begin{algorithmic}[1]
\Require Traffic loads of SBSs $\lambda_{i,j}$, maximum number of layers $L$
\Ensure Clustered SBSs with estimated traffic loads

\Procedure{MLC\_k\_means}{$\lambda$, $L$}
\State Determine the optimal number of clusters $k$ using the elbow method
    \State Initialize layer count $l = 1$
    \While{$l \leq L$}
        \State Perform $k$-means clustering on $\lambda$ to form $k$ clusters
        \algdef{SE}[FOR]{For}{EndFor}[1]{\textbf{for}\ #1\ \algorithmicdo}{\textbf{end for}}%
        \For{each cluster $C_k$}
        \State Calculate the mean traffic load $\mu_\text{m}$
            \For{each sleeping SBS in $C_k$}
                \State Estimate the traffic load as $\mu_\text{m}$
            \EndFor
        \EndFor
        \State Update $\lambda$ with estimated ones for sleeping SBSs
        \State Increment the layer count $l$ by 1
    \EndWhile
    \State \Return The final clusters with estimated traffic loads
\EndProcedure
\end{algorithmic}
\label{alg1}
\end{algorithm}

\subsection{Geographical Distance-Based Traffic Load Estimation}
This method considers the proximity of neighboring cells to estimate the traffic load of a sleeping SBS. It includes two sub-methods based on the presence of a weighting mechanism: distance-based without and with weighting.
 
\subsubsection{Distance-Based without Weighting} In this approach, the traffic load of a sleeping SBS is estimated by averaging the traffic loads of its neighboring cells, arranged incrementally based on proximity. All neighboring cells contribute equally to the estimation, regardless of their distance. The estimated traffic load, $\hat\lambda _i$, is calculated as
\begin{equation}
    \hat\lambda _i  = \frac{1}{N}\sum\limits_{j = 1}^N {\lambda _j, } 
\label{eq7}
\end{equation}
where ${{\lambda _j }}$ represents the traffic load of the $j$-th neighboring cell, and $N$ is defined as the number of neighboring cells included in the estimation process.

\subsubsection{Distance-Based with Weighting} This method refines the previous approach by assigning different weights to neighboring cells based on their distance from the sleeping SBS. The closer a cell is, the more it influences the estimated traffic load. The weighted traffic load, $\hat\lambda _i$, is calculated as
\begin{equation}
\hat \lambda _i  = \frac{{\sum\limits_{j = 1}^N {\lambda _j  \times w_{i,j} } }}{{\sum\limits_{j = 1}^N {w_{i,j} } }}.
\label{eq8}
\end{equation}
The weighting factor, ${w_{i,j}}$, is defined as
\begin{equation} 
w_{i,j}  = \frac{{d_{\max } }}{{d_{i,j}^n }},\;\;\;\;
n \in \{ 1,2,...,k\} 
\label{eq9}
\end{equation}
where ${d_{\max }}$ is the maximum distance between the sleeping SBS and its neighboring cells included in the estimation, and ${d_{i,j}}$ is the distance between the sleeping SBS ${i}$ and the neighboring SBS ${j}$.
\subsection{Random Cell Selection Traffic Load Estimation}
This approach utilizes a random selection of surrounding cells for traffic load estimation. It comprises two variations: random selection without weighting and random selection with weighting.
\subsubsection{Random Selection without Weighting}
The traffic load of a sleeping SBS is estimated based on the average traffic load of randomly selected surrounding SBSs. The estimation formula is similar to \eqref{eq7}, but the selection of ${j}$ is random.
\subsubsection{Random Selection with Weighting}
This variation applies a weighting mechanism to the randomly selected surrounding cells, same as \eqref{eq9}, where ${j}$ is selected randomly. The weighting factors vary based on the distance to the sleeping SBS, enhancing the accuracy of the traffic load estimation.

\section{Performance Evaluation}\label{sec:performance}
In this section, we assess the efficacy of our proposed methods for estimating the traffic load of sleeping SBSs. 
The evaluation is conducted using the Milan data set, as described earlier in Section II-C. 
The parameters used for simulations are outlined in Table \ref{table:nonl}. Our data collection process involves averaging the traffic for each SBS at each time slot over a month, followed by applying the proposed algorithms with a randomized selection of SBSs for each iteration. The primary metric for evaluating the performance of our methods is the traffic load estimation error, defined as the ratio of the difference between the actual and estimated traffic load to the actual traffic load. \\
\begin{centering}
\begin{table}[h!] 
\caption{SIMULATION PARAMETERS}
\centering 
\begin{tabular}{|c|c|} 
  \hline 
Parameters&Value
\\ 
\hline
Number of SBSs&5000
\\
Number of time slots&144
\\
Time slot duration&10 m
\\
Number of days&30
\\
Number of iteration&300
\\
Optimal $K$ using elbow method&3
\\
\hline
    \end{tabular} \label{table:nonl} 
    \vspace{-2em}
    \end{table}
    \end{centering}

\begin{figure}[t]
\centerline{\includegraphics[width = 9.cm ]{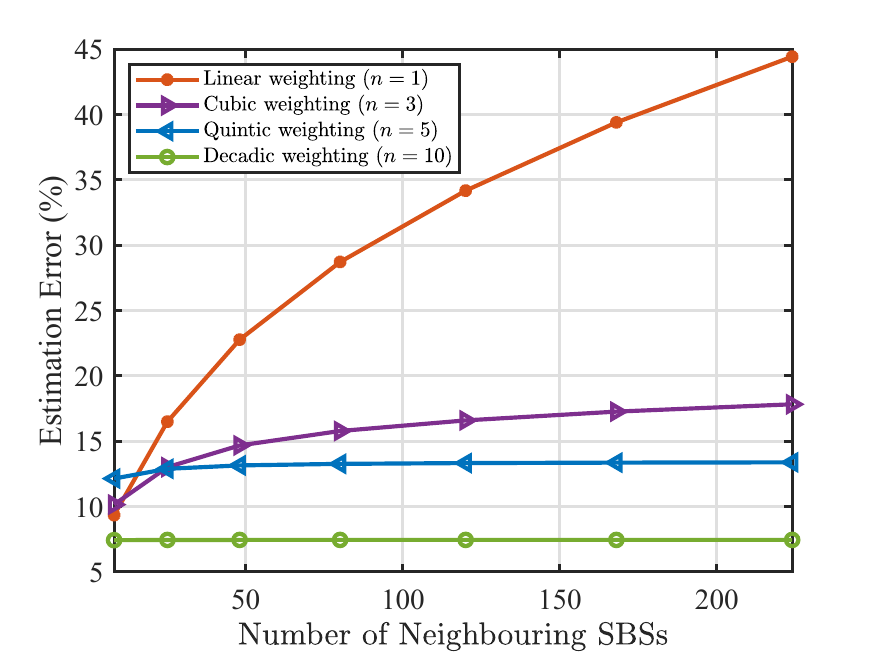}}
\caption{Estimation error for distance-based cell selection with weighting.}
\label{fig-2}
 
\end{figure}

\begin{figure}[t]
\centerline{\includegraphics[width = 9.cm ]{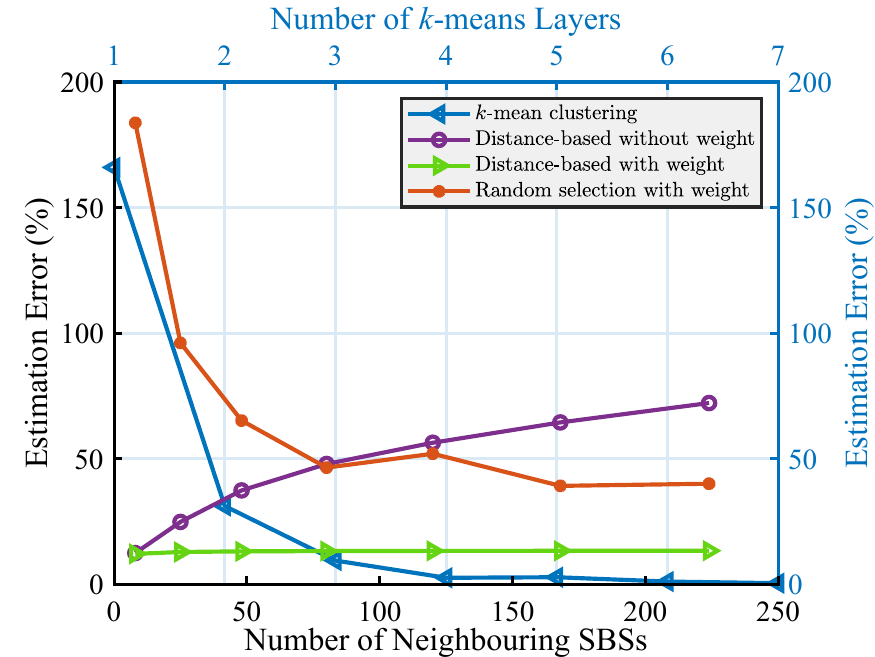}}
\caption{The estimation error for different methods. Two $x$- and two $y$-axes are considered: blue one is for clustering based while black one is for the rest of the approaches.}
\label{fig-3}
\vspace{-1em}
\end{figure}

\indent Fig. \ref{fig-2} presents the relationship between traffic load estimation error and the number of neighboring SBSs for the distance-based estimation method with weighting. The results demonstrate how the estimation error is affected by varying the power of distance, $n$, in \eqref{eq9}, with values of 1, 3, 5, and 10. Two key observations emerge from this analysis:
\begin{itemize}
\item An increase in the number of neighboring SBSs leads to a decrease in the accuracy of the traffic load estimation. This trend suggests that incorporating too many neighboring SBSs diminishes the relevance of each individual cell's data.

\item The performance of the distance-based method with weighting improves as $n$ increases. This improvement is attributable to the method's emphasis on the proximity of neighboring cells; closer neighbors have a more significant impact on the estimation than those farther away. For example, the estimation error rises to 45\% with 
$n=1$ for a large number of neighboring SBSs, but it drops below 15\% for $n=5$. This trend is even more pronounced for $n=10$, where the estimation error appears less dependent on the number of contributing neighbors.
\end{itemize}
Fig. \ref{fig-3} compares the performance of all three proposed methods in terms of the load estimation error. Given the different nature of the variable in $k$-means clustering compared to the other methods, Fig. \ref{fig-3} features two $x$-axes. As the number of layers in $k$-means clustering increases, indicating a more homogeneous traffic load within each cluster, the estimation error decreases. The results show that it is possible to achieve near 0\% estimation error with seven layers of clustering. The trend for the distance-based method without weighting is similar to the weighted case, but the estimation error increases more rapidly with the number of neighboring cells, as distance is not a factor in the estimation.
Conversely, the random selection method with weighting shows a decreasing trend in estimation error. This improvement is attributed to the increased number of contributing SBSs and the emphasis on closer neighbors, enhancing the accuracy of the traffic load estimation.
The results for the random selection without weighting are not presented, as the estimation errors were found to be high and fluctuated significantly due to the random nature of neighbor selection and the lack of distance consideration.

\begin{figure}[t]
\centerline{\includegraphics[width = 9.cm ]{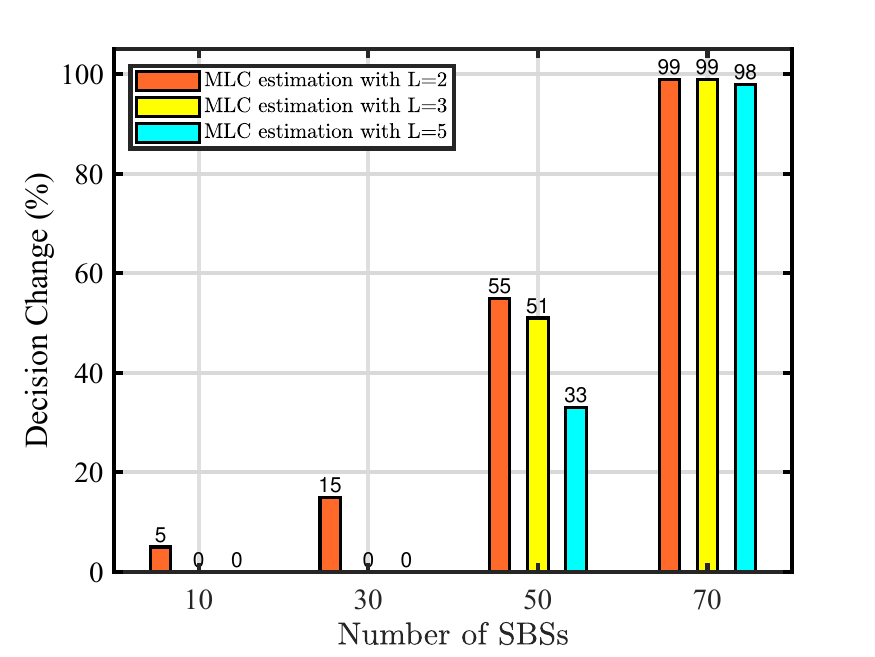}}
\caption{Decision change for different number of layers in MLC.}
\label{fig-4}
\end{figure}

Fig. \ref{fig-4} illustrates the discrepancies between the actual and estimated state vectors in the context of $k$-means clustering with varying numbers of layers. 
The results in Fig.~\ref{fig-4} demonstrate that an increase in the number of SBSs leads to greater variation in the state vector. 
Further, for a given number of SBSs, augmenting the number of layers in the MLC process results in a reduction of this discrepancy, which can be attributed to the inverse relationship between estimation error and the number of layers in $k$-means clustering, as observed in Fig.~\ref{fig-3}. This discrepancy is particularly pronounced when there are fewer SBSs in the vHetNet but diminishes as the number of SBSs increases. In a densely populated vHetNet with 70 SBSs, the difference becomes negligible, leading to a decision change rate of 98\%-99\%.

\begin{figure}[h!]
\centerline{\includegraphics[width = 9.cm ]{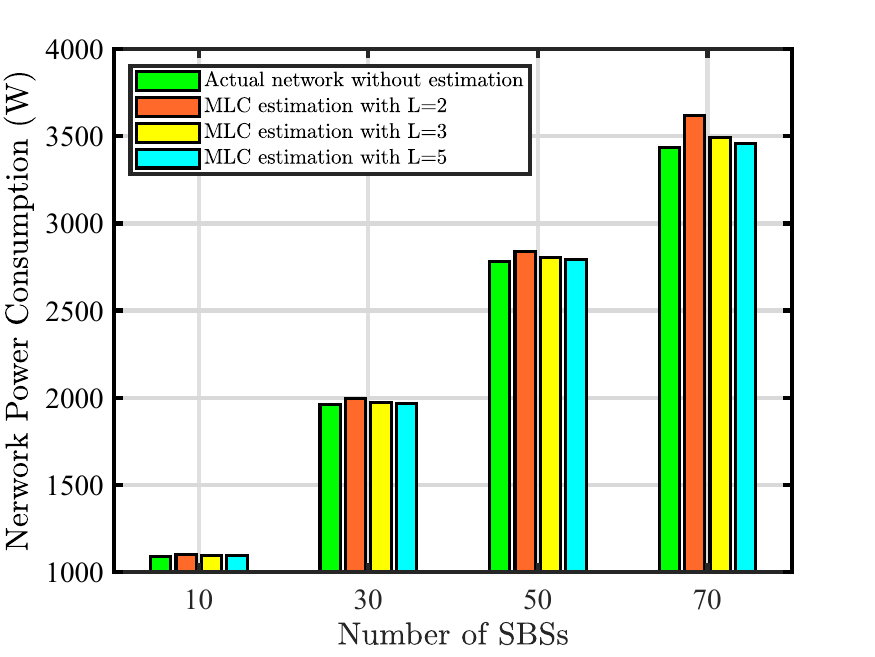}}
\caption{Network power consumption for different number of layers in MLC.}
\label{fig-5}
\end{figure}

Fig.~\ref{fig-5} depicts the network power consumption as a function of the number of SBSs, $s$, comparing the results for actual versus estimated traffic loads of sleeping SBSs. 
The estimation scenario, as described in Fig. \ref{fig-4}, is conducted using the $k$-means clustering algorithm with varying numbers of layers, $L$. 
Consistent with expectations, the results demonstrate that network power consumption increases with $s$, in both actual and estimated scenarios. Another key observation is that, for any given $s$, an increase in $L$ results in the estimated network power consumption more closely aligning with the actual consumption. 
This convergence is attributed to the enhanced accuracy of the traffic load estimation with higher layers in the MLC process. 
Interestingly, despite substantial changes in the state vector for larger $s$, the results indicate no significant difference in network power consumption. This observation can be explained by the efficacy of the offloading algorithm proposed in~\cite{10304250}, which effectively manages the network power consumption even with changes in the state vector.

It is important to acknowledge that the results presented in this study are specific to the Milan dataset. While different datasets may yield different outcomes, we intuitively believe that the findings offer generic insights and trends applicable to the developed solutions. Such an acknowledgment underscores the relevance of our methodologies across varying network scenarios, despite the inherent dataset-specific limitations.
\section{Conclusion}\label{sec:conclusion}
This research tackles a significant challenge in vHetNets—accurately estimating traffic loads for SBSs in sleep mode. The lack of reliable data on the traffic load of sleeping SBSs poses a critical obstacle, affecting the effectiveness of power consumption optimization strategies in vHetNets. By focusing on the cell load estimation problem, our study highlights its significance and the pressing need for effective solutions.
To address this issue, we introduced three innovative spatial interpolation methods: clustering-based, distance-based, and random neighboring selection, which aim to fill the data gap and improve power consumption management in vHetNets. 
Using the Milan data set for validation, our analysis confirms the effectiveness of these methods in accurately estimating traffic loads. This accuracy is essential for implementing power optimization techniques, ensuring that energy-efficient strategies are both feasible and practical.

\ifCLASSOPTIONcaptionsoff
  \newpage
\fi
\bibliographystyle{IEEEtran}
\bibliography{IEEEabrv,Bibliography}
\vfill
\end{document}